\documentclass{article}
\usepackage{spconf,amsmath,graphicx}
\usepackage{url}
\usepackage{amsmath}
\usepackage{amssymb}    

\usepackage{booktabs} 
\usepackage{siunitx}
\usepackage{multirow}
\usepackage{todonotes}
\usepackage{adjustbox}

\usepackage{enumitem}
\setlist{nosep, leftmargin=14pt}

\usepackage{mwe} 

\newcommand{\s}[1]{\mathcal{#1}} 
\newcommand{\sD}{\s{D}}  
\DeclareMathOperator*{\argmax}{arg\,max}

\title{Useful Nonrobust features are ubiquitous in biomedical images}
\name{%
\begin{tabular}{c}
\itshape Coenraad Mouton$^{\star\dagger}$ \qquad Randle Rabe$^{\dagger}$ \qquad Niklas C Koser$^{\star}$ \\
\itshape Nicolai Krekiehn$^{\star}$ \qquad Christopher Hansen$^{\star}$ \qquad Jan-Bernd Hövener$^{\star}$ \qquad Claus-C. Glüer$^{\star}$
\end{tabular}}
\address{$^{\star}$ Section Biomedical Imaging, Department of Radiology and Neuroradiology, \\ University Hospital Schleswig-Holstein, Kiel University \\
    $^{\dagger}$ Faculty of Engineering, North-West University, South Africa}
\begin{document}
%
\maketitle
%
\begin{abstract}
We study whether deep networks for medical imaging learn useful nonrobust features - predictive input patterns that are not human interpretable and highly susceptible to small adversarial perturbations - and how these features impact test performance. We show that models trained only on nonrobust features achieve well-above-chance accuracy across five MedMNIST classification tasks, confirming their predictive value in-distribution. Conversely, adversarially trained models that primarily rely on robust features sacrifice in-distribution accuracy but yield markedly better performance under controlled distribution shifts (MedMNIST-C). Overall, nonrobust features boost standard accuracy yet degrade out-of-distribution performance, revealing a practical robustness-accuracy trade-off in medical imaging classification tasks that should be tailored to the requirements of the deployment setting.\footnote{This manuscript is an extended version of a paper presented at the IEEE International Symposium on Biomedical Imaging (ISBI), 2026.}
\end{abstract}
\section{Introduction}
\label{sec:intro}
\begin{figure}[!t]
  \centering
  \includegraphics[width=0.9\columnwidth]{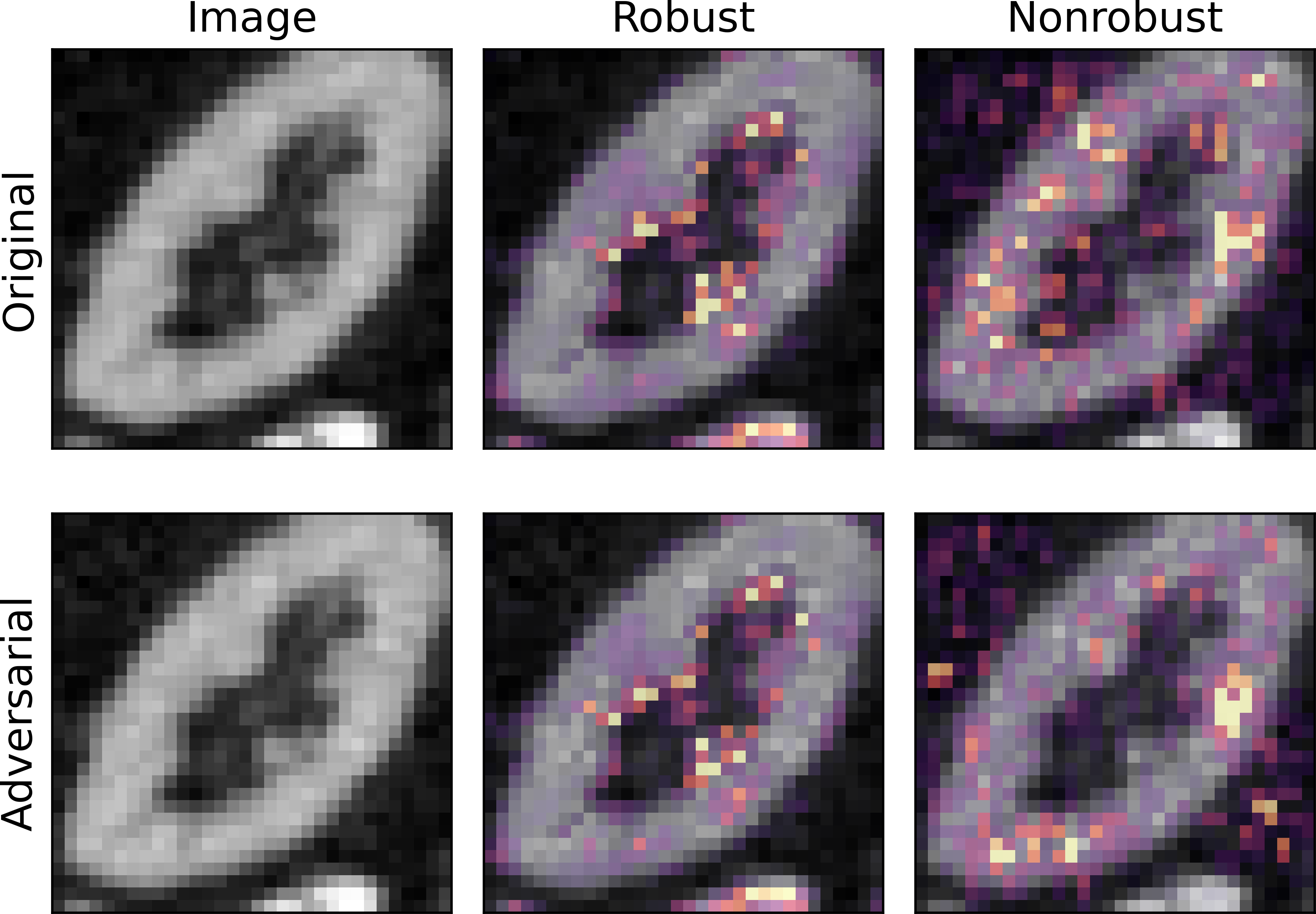}
    \caption{Saliency map comparison between robust and nonrobust organ-classification models for a right-kidney CT slice (top) and a slightly perturbed adversarial version (bottom). The robust model’s attributions remain stable between the original and adversarial images and concentrate on kidney-relevant anatomy, whereas the nonrobust model’s attributions change markedly under perturbation and appear almost noise-like. Overlays are normalized to $[0,1]$; brighter colors indicate larger attribution.}
  \label{fig:feature_attribution}
\end{figure}
Deep neural networks (DNNs) are increasingly being deployed in medical imaging domains such as radiology, digital pathology, and ophthalmology. Given the high-stakes of these application domains, it is critical that these models are reliable, trustworthy and (ideally) explainable. Despite this, prior work on natural images has shown that neural networks learn \emph{nonrobust features}: input patterns which are extremely susceptible to small perturbations and completely uninterpretable to humans, yet highly predictive~\cite{ilyas_nr,nr_ntk}. It is further shown that the existence of adversarial examples -  tiny, virtually imperceptible perturbations to images that reliably cause misclassifications - can be ascribed (at least in part) to a model's reliance on nonrobust features (see Fig.~\ref{fig:feature_attribution}). 
Such features are naturally troublesome should they also occur in the medical domain: not only would they hinder any attempts at model interpretation~\cite{robustness_at_odds} and increase susceptibility to adversarial examples, but one must then also consider whether a user should \emph{trust} a prediction that was made based on uninterpretable and fragile patterns.

To approach this matter, we consider two fundamental questions: (1) Do DNNs learn useful, well-generalizing nonrobust features from various biomedical imaging modalities, and (2) How essential are such nonrobust features for high performance.

Accordingly, we consider the role that robust and nonrobust features play in several classification tasks spanning diverse imaging modalities such as computed tomography (CT), radiography,  ultrasound, and histopathology images. Furthermore, we conduct our investigation in both an in-distribution and an out-of-distribution (OOD) setting.  

Our primary contribution is that we provide the first systematic investigation of the generalization properties of robust and nonrobust features in medical image classification tasks.  


\section{Robust and nonrobust features}
In this section we introduce key notation and formalize robust and nonrobust features.
Let $f$ denote a deep neural network (DNN) with parameters $\mathbf{\theta}$ trained on a dataset of sample-label pairs $\sD = \{(\mathbf{x}_i, y_i)\}_{i=1}^N$ using a suitable loss function $\mathcal{L}$. We define a \emph{useful feature} learned by $f$ for a class $c$ as a function $g_c: \mathbb{R}^d \rightarrow \mathbb{R}$ which is correlated with $y=c$ under $\sD$. 
Intuitively, a feature can be considered a single learned property or concept of the input data, for example `cat ears' or `human face' which aids in classification for a given class $c$.

We then further distinguish between \emph{robust} and \emph{nonrobust} features. Consider an adversarial perturbation $\boldsymbol{\delta}$ such that
\begin{equation}
\label{eq:delta_perturbation}
\boldsymbol{\delta}^{\star}
= \argmax_{\lVert \boldsymbol{\delta}\rVert_{\infty}\le \epsilon}
\mathcal{L}\left(y, f_{\theta}(\mathbf{x}+\boldsymbol{\delta})\right)
\end{equation}
where $\epsilon$ is some small perturbation bound, for example $\frac{4}{255}$ is commonly used for image data~\cite{robustbench}.
A feature $g_c$ is then defined as \emph{robust} if $g_c(\mathbf{x}+\boldsymbol{\delta}^\star)$ remains correlated with $c$, but \emph{nonrobust} if $g_c(\mathbf{x}+\boldsymbol{\delta}^\star)$ is instead correlated with some $j\neq c$. See \cite{ilyas_nr} for formal definitions.

More intuitively, we aim to convey that features are robust if they remain readily recognizable and correlated with a class label after some small adversarial perturbation. For example, a human face is still easily identifiable after slight distortion of the underlying pixel values. Conversely, nonrobust features are easily switched from being correlated with one class to another by tiny perturbations. A side-effect of this brittle nature is that these features are generally not recognizable, or interpretable, by humans, and therefore appear as noise-like patterns when isolated in the image domain. 

To illustrate, Fig.~\ref{fig:feature_attribution} compares the saliency (gradient of the predicted class score w.r.t. the input) of an adversarially robust and an adversarially nonrobust classification model trained on OrganSMNIST~\cite{medmnist}. This is done for a test image and an adversarially perturbed version that is misclassified by the nonrobust model. The visual contrast between these saliency maps mirrors our definitions: robust features are stable and more anatomically meaningful, whereas nonrobust features are brittle and easily altered by small perturbations. 

\section{Experimental Setup}
In this section we describe the datasets, models, and procedures used to isolate and evaluate robust vs. nonrobust features.
\begin{figure*}[t]
  \centering
  \includegraphics[width=0.70\textwidth]{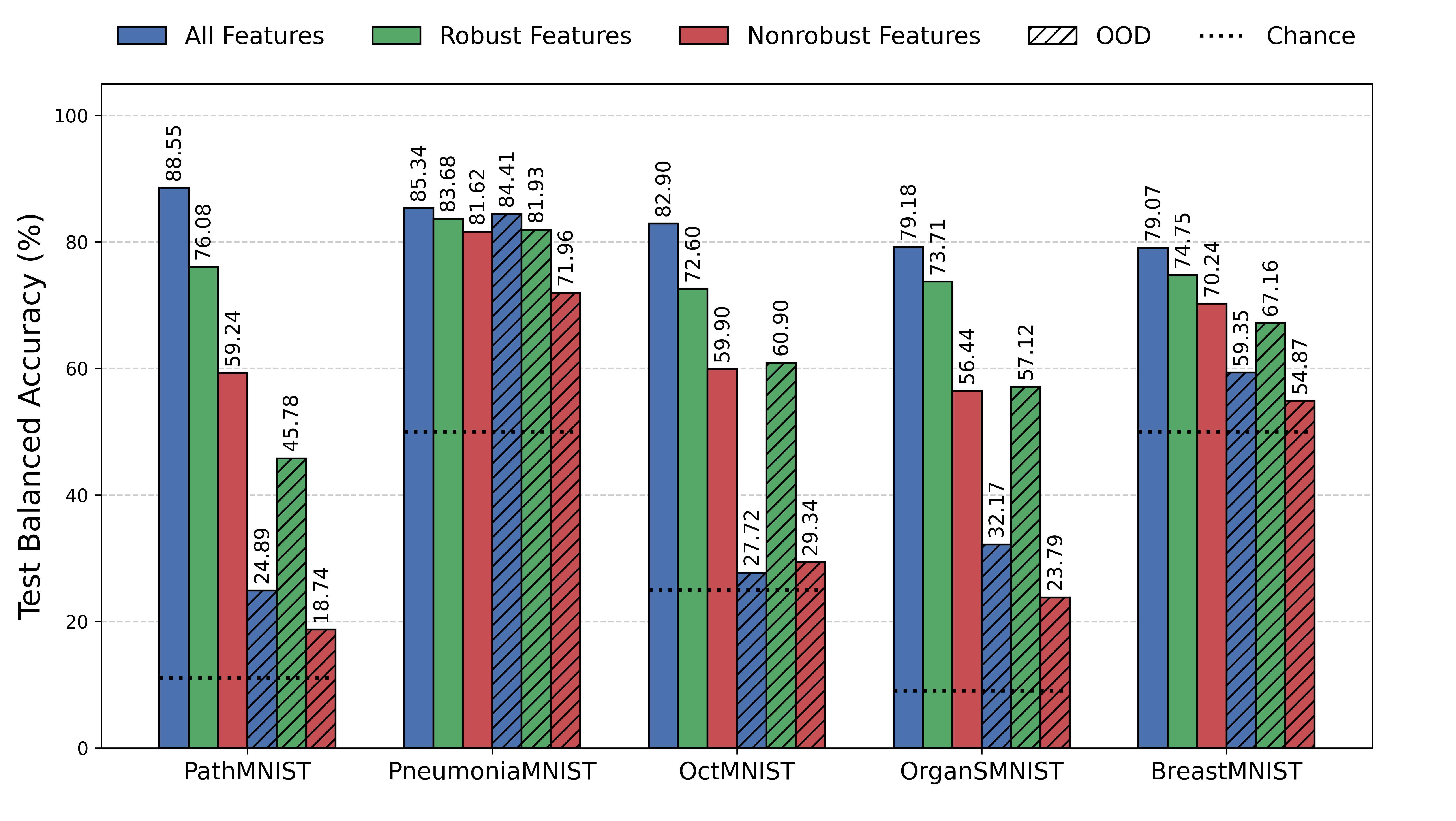}
  \caption{Test 
  accuracy comparison of DNNs using different features for $5$ MedMNIST datasets. Solid bars: in-distribution performance. Dashed bars: average out-of-distribution performance. }
  \label{fig:nonrobust-all}
\end{figure*}
\subsection{Isolating nonrobust features}
Our first avenue of inquiry concerns determining whether well generalizing nonrobust features can be found in the various medical imaging datasets we consider. More specifically, we aim to establish whether nonrobust features alone suffice for non-trivial performance (better than random guessing) on these classification tasks. 

To investigate this, we borrow an experimental setup from Ilyas et al.~\cite{ilyas_nr} and proceed as follows. Consider a medical imaging dataset $\sD$, split into two subsets, $\sD^{\text{train}}$ and $\sD^{\text{test}}$. Furthermore, let $f$ correspond to a DNN trained using $\sD^{\text{train}}$ such that it performs well on $\sD^{\text{test}}$. This model should presumably rely on both robust and (should they exist) nonrobust features in order to make predictions. 

Our goal is to construct a new dataset, $\hat{\sD}^{\text{train}}$, where the only correlation between the labels and inputs is the nonrobust features learned by $f$. We construct such a dataset as follows. Consider all $m$ samples $(\mathbf{x}, y) \in \sD^{\text{train}}$ that are correctly classified by $f$ such that $f(\mathbf{x})=y$. For each of these samples, we randomly select a target class $t$ and then create an adversarial example $\hat{\mathbf{x}} = \mathbf{x} + \boldsymbol{\delta}$ (per equation~\ref{eq:delta_perturbation}) such that $f(\hat{\mathbf{x}}) = t$. The dataset $\hat{\sD}^{\text{train}}$ is then constructed from these adversarial examples and labels, i.e. $\hat{\sD}^{\text{train}} = \bigl\{\,(\hat{\mathbf{x}}_i,\; t_i)\bigr\}_{i=1}^{m}$. We then train a new classifier on $\hat{\sD}^{\text{train}}$ and evaluate it on the original, unadulterated test set $\sD^{\text{test}}$. This allows us to determine to which extent the nonrobust features learned by $f$ alone can suffice for classification. 

More simply, we construct a new training set consisting solely of adversarial examples and assign each example the label of its randomly chosen adversarial target class. Intuitively, this procedure switches the image’s nonrobust features from the source class to the target class (as learned by the original classifier) while leaving the robust, more human-interpretable features largely intact. Therefore, the robust features are now uncorrelated with the new label, and the entire dataset appears (to a trained radiologist or specialist) unchanged but mislabeled. The only systematic signal that remains aligned with the (new) train set labels are the switched nonrobust features. Consequently, if a model trained on this dataset achieves non-trivial (better than chance) accuracy on the original test set with the original labels, it must be doing so by exploiting nonrobust features alone.

We perform this experiment using $5$ medical imaging classification datasets of distinct imaging modalities from the MedMNIST~\cite{medmnist} collection. For each dataset, we (1) train a WRN-16-8 (WideResNet~\cite{wide_resnet}) model on the original dataset, (2) use this trained model to construct a nonrobust feature dataset, then (3) train a new WRN-16-8 model on the constructed dataset, and finally (4) evaluate both models on the original test set. We perform an extensive hyperparameter search for both steps (1) and (3) to ensure that each model is well optimized (see Appendix~\ref{sec:app_hp}). Furthermore, we use $32\times32$ sized images in all cases (resized from the original $64 \times 64$) with the predefined train/val/test splits, and we explicitly do not use any pretrained weights, as this would introduce additional features. Finally, we rely on a $\ell_{\infty}$ perturbation bound of $\epsilon =  \frac{4}{255}$ for step (2), where Equation~\ref{eq:delta_perturbation} is approximated using 100-step PGD (projected gradient descent)~\cite{madry_at}. 
\subsection{Isolating robust features}
We now ask the converse question:  How well do models that rely primarily on \emph{robust} features generalize? To this end, we make use of adversarial training~\cite{madry_at}, implemented via the TRADES~\cite{trades} loss function:
\begin{equation}
\label{eq:trades_loss}
\adjustbox{max width=\columnwidth}{$
\mathcal{L}_{\text{TR}}(\mathbf{x},y;\theta)
= \ell_{\text{CE}}\bigl(y,f_{\theta}(\mathbf{x})\bigr)
+\beta\, \mathrm{KL}\!\bigl( f_{\theta}(\mathbf{x}) \,\|\, f_{\theta}(\mathbf{x}+\boldsymbol{\delta}^{\star}) \bigr)
$}
\end{equation}
where $ \ell_{\text{CE}}$ is the cross-entropy loss, $\mathrm{KL}$ is the Kullback-Leibler divergence, and we assume here that the output of $f_{\theta}$ is a probability vector (not logits or a scalar class prediction). 
The TRADES loss essentially balances the performance on clean data (left term) with the performance on adversarial data (right term) according to a hyperparameter $\beta$. This encourages the model to ignore nonrobust cues, and allows us to determine how performance changes as the reliance on nonrobust features decreases. 

During training, we incorporate most of the modern techniques that assist adversarial training, such as label smoothing, exponential weight averaging, swish activations, and cosine annealing\footnote{We also use these methods when training the base and nonrobust feature models, to ensure they are comparable.}, while early stopping is performed on an adversarial validation set~\cite{robust_overfitting, wang_at_setup, bag_of_tricks}. Furthermore, the adversarial perturbations are crafted using 10-step PGD with $\epsilon=\frac{4}{255}$, and we set $\beta=5$. As previously done, we perform an extensive hyperparameter search per model to ensure that it is well optimized. 
\subsection{Evaluating OOD performance}
Nonrobust features are defined as `nonrobust' because they are extremely susceptible to tiny \emph{adversarial} perturbations. However, this does not necessarily imply that these features are brittle to other distribution shifts. To determine this, we compare the degree to which the various models are sensitive to controlled corruptions using the MedMNIST-C test sets~\cite{medmnist_corrupted}. These consist of the original MedMNIST test set altered by various transformations, such as contrast adjustments, Gaussian noise, blurring, JPEG compression, and others. We report on the average performance of each model across these different distribution shifts. 

\subsection{Evaluating adversarial performance}
\label{sec:autoattack-explanation}

As a final experiment, we verify that our adversarially trained models are truly more robust to adversarial attacks. We measure this using the strong ensemble method AutoAttack~\cite{auto_attack} with $\epsilon = \frac{4}{255}$ perturbations.

\section{Results}

In Fig.~\ref{fig:nonrobust-all} we report the in-distribution (solid bars) and OOD performance (dashed bars) across five MedMNIST datasets, comparing base models (trained on all features; blue), robust models (green), and nonrobust-only models (red). Balanced accuracy is used throughout due to class imbalance.
\subsection{Comparing in-distribution test performance}
When considering the in-distribution results, we observe that in all cases the nonrobust-only models achieve well above chance performance (indicated by the dashed horizontal lines). For example, the nonrobust-only model achieves a balanced accuracy of $59.24\%$ and $81.62\%$ on PathMNIST and PneumoniaMNIST, respectively, which is substantially higher than the $11\%$ and $50\%$ that would be achieved by random guessing.  This is a strong indication that useful nonrobust features are present and learnable in these datasets.

It is also clear from these results that using all available features consistently yields the best test performance; relying primarily on robust features is second-best; and nonrobust-only performs worst. Furthermore, the performance difference between using all available features and only robust features can be significant, e.g. $88.55\%$ versus $76.08\%$ balanced accuracy for PathMNIST, and similarly $82.90\%$ versus $72.60\%$ for OctMNIST. The implication of this is that the addition of nonrobust features significantly improves performance in the standard classification setting. 
\subsection{Comparing OOD test performance}
With respect to OOD performance, we find that adversarially trained models that primarily rely on robust features provide the best performance (except PneumoniaMNIST). Compared to using all features, the difference can be substantial, e.g. the balanced accuracy increases from $24.89\%$ to $45.78\%$ for PathMNIST and from $27.72\%$ to $60.90\%$ for OctMNIST when the model is adversarially trained.  

By contrast, the nonrobust-only models generally mirror the OOD performance drop of the base models, which is often very large. We believe that in both cases this decrease in performance is tied to the models' reliance on nonrobust features. Despite this, we note that in all cases we still observe (marginally) higher than chance performance on the OOD test sets, and in one case strong performance (PneumoniaMNIST).

\subsection{Comparing adversarial test performance}
\label{sec:adversarial_performance}

We find that all base and nonrobust-only models are highly susceptible to adversarial attacks, with an adversarial accuracy of $\leq 0.04\%$. Conversely, the adversarially trained models achieve between $57\%$ and $74\%$ depending on the dataset, confirming that they are (unsurprisingly) much more adversarially robust. These results are omitted from Fig.~\ref{fig:nonrobust-all} but can be found in Appendix~\ref{sec:app_autoattack}.
\section{Discussion}

Let us consider our various observations and revisit the two originally posed questions:
\begin{enumerate}
    \item \textbf{Do DNNs learn useful nonrobust features from medical images?} Yes - nonrobust features alone yield accurate classification on several datasets. While still inferior to robust features, they perform well above chance. By contrast, under OOD shifts, accuracy drops to only marginally above chance, and to zero under adversarial attacks.
    \item \textbf{How essential are nonrobust features?} In-distribution, discouraging nonrobust features reduces classification performance (by $2-12\%$). Conversely, in the OOD setting, the presence of nonrobust features alongside robust ones lowers accuracy on most datasets (by $8-33\%$), whereas robust features alone are much more invariant to distribution shifts and adversarial attacks.
\end{enumerate}
These answers expose an uncomfortable truth: despite nonrobust features being uninterpretable and extremely vulnerable to both adversarial perturbations and distribution shifts, they are generally very useful in the standard classification setting. This conclusion echoes prior work on natural images which note a trade-off between adversarial robustness and accuracy~\cite{robustness_at_odds,trades}. It is therefore difficult to determine whether the use of nonrobust features should be encouraged or dissuaded. We believe it depends on the expected deployment setting: if certain distribution shifts are expected and model safety and interpretation is paramount, nonrobust features are to be avoided. Conversely, if absolute in-distribution performance is the goal, they are likely best left included.

Besides this performance assessment, it is also crucial to consider how nonrobust features affect the trustworthiness of an imaging model. i.e. whether a user (e.g. a physician) can reasonably rely on the mechanism behind a prediction. Nonrobust features likely reduce trustworthiness: they are generally uninterpretable and fragile to small, plausible changes, making decisions harder to justify. That said, it is unclear whether users would prefer a more interpretable, shift-robust model over a more accurate one. A promising avenue for future work is to assess, in realistic clinical settings, which trade-off clinicians actually prefer for different use cases.  

Despite our strong empirical results, we must also note several limitations. While we have compared OOD performance, we focus on artificial corruptions and believe that an evaluation on natural distribution shifts (e.g. independently sourced data) would also be insightful. Similarly, we did not explicitly train models to be robust to the MedMNIST-C shifts (e.g. with data augmentation), which could potentially shrink the gap between robust and nonrobust features. Furthermore, we solely focus on CNNs (WRN-16-8) trained on two-dimensional low resolution images. Although we expect similar trends for modern vision transformers and higher-resolution or three-dimensional imaging, this remains to be verified. Finally, we confine our study to the task of classification and we believe that an extension to segmentation or detection settings would also prove valuable.

\section{Conclusion}

In conclusion, we have shown that (1) nonrobust features are present in medical imaging datasets and exploited by DNNs, (2) these features are predictive in-distribution, and suppressing them harms in-distribution accuracy, yet (3) their presence degrades OOD performance, whereas robust features alone are markedly more invariant. This clarifies a robustness-accuracy trade-off that should be taken into account.
\newpage
\section{COMPLIANCE WITH ETHICAL STANDARDS}
This research study was conducted retrospectively using human subject data made available in open access by the MedMNIST and MedMNIST-C datasets~\cite{medmnist,medmnist_corrupted}. Ethical approval was not required as confirmed by the license attached with the open access data.
\section{ACKNOWLEDGEMENTS}
This project was supported by the modular AI Imaging Pipelines (mAIPipes) Grant, Application No. 22024025 KI-Förderrichtlinie Schleswig-Holstein, Germany, and supported in part by the National Research Foundation of South Africa (Ref Numbers PSTD23042898868, RCDL240215206999).
\bibliographystyle{IEEEbib}
\bibliography{strings,refs}
\clearpage
\appendix
\onecolumn
\section{Additional results}

\subsection{Area under the curve}
\label{sec:app_auc}

\begin{figure}[h]
  \centering
  \includegraphics[width=0.70\textwidth]{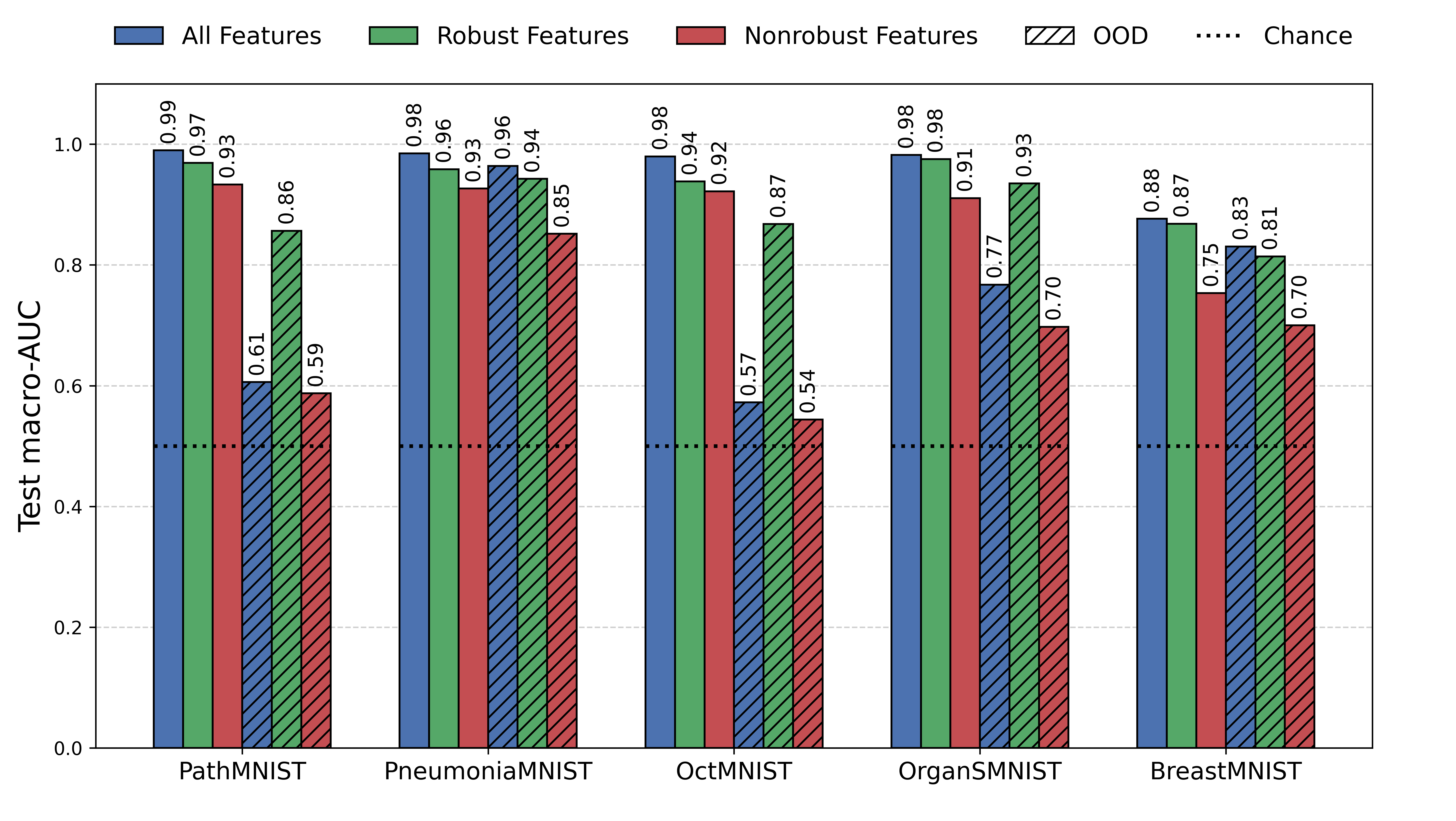}
  \caption{The same as Fig.~\ref{fig:nonrobust-all} using AUC as the performance metric.}
  \label{fig:nonrobust-all-auc}
\end{figure}

In the main paper, we rely on balanced accuracy to compare model performance across datasets. For completeness, we also report results using the area under the ROC curve (AUC) as it is a commonly used metric in medical imaging. Fig.~\ref{fig:nonrobust-all-auc} presents the same experiments as Fig.~\ref{fig:nonrobust-all}, but evaluated using macro-averaged one-vs-rest AUC. The overall trends are consistent with those observed for balanced accuracy.

\subsection{Detailed adversarial performance}
\label{sec:app_autoattack}
As mentioned in the main paper, we evaluate adversarial robustness using AutoAttack with $\epsilon=\frac{4}{255}$. Models that rely on nonrobust features are almost completely broken under these attacks, achieving $0\%$ balanced accuracy across datasets.\footnote{With the exception of the ``All Features'' model on BreastMNIST, which correctly classifies a single sample under attack.} By contrast, adversarially trained models retain substantial performance, achieving between $57\%$ and $74\%$ balanced accuracy. Table~\ref{tab:autoattack_robust} compares the adversarial performance of these models to the clean performance.

\begin{table}[h]
\centering
\caption{Clean and AutoAttack balanced accuracy (\%) for the adversarially trained robust models.}
\label{tab:autoattack_robust}
\begin{tabular}{lcc}
\toprule
Dataset & Clean Balanced Accuracy (\%) & AutoAttack Balanced Accuracy (\%) \\
\midrule

PathMNIST      & 76.08 & 57.45 \\
PneumoniaMNIST & 83.68 & 73.85 \\
OctMNIST       & 72.60 & 59.00 \\
OrganSMNIST    & 73.71 & 67.88 \\
BreastMNIST    & 74.75 & 64.79 \\
\bottomrule
\end{tabular}
\end{table}   
\subsection{The effect of $\epsilon$}
Throughout the main paper, we use a maximum $\ell_\infty$ perturbation of $\epsilon=\frac{4}{255}$ to define robust and nonrobust features. This choice affects both the construction of the nonrobust-only datasets and the adversarial training procedure used to obtain robust models, as both are effectively trained using $\epsilon=\frac{4}{255}$ perturbations. Here, we examine whether our conclusions change when using a larger perturbation bound of $\epsilon=\frac{8}{255}$, which is commonly used in natural image benchmarks such as CIFAR-10~\cite{robustbench}.
 
\begin{figure}[h]
  \centering
  \includegraphics[width=0.70\textwidth]{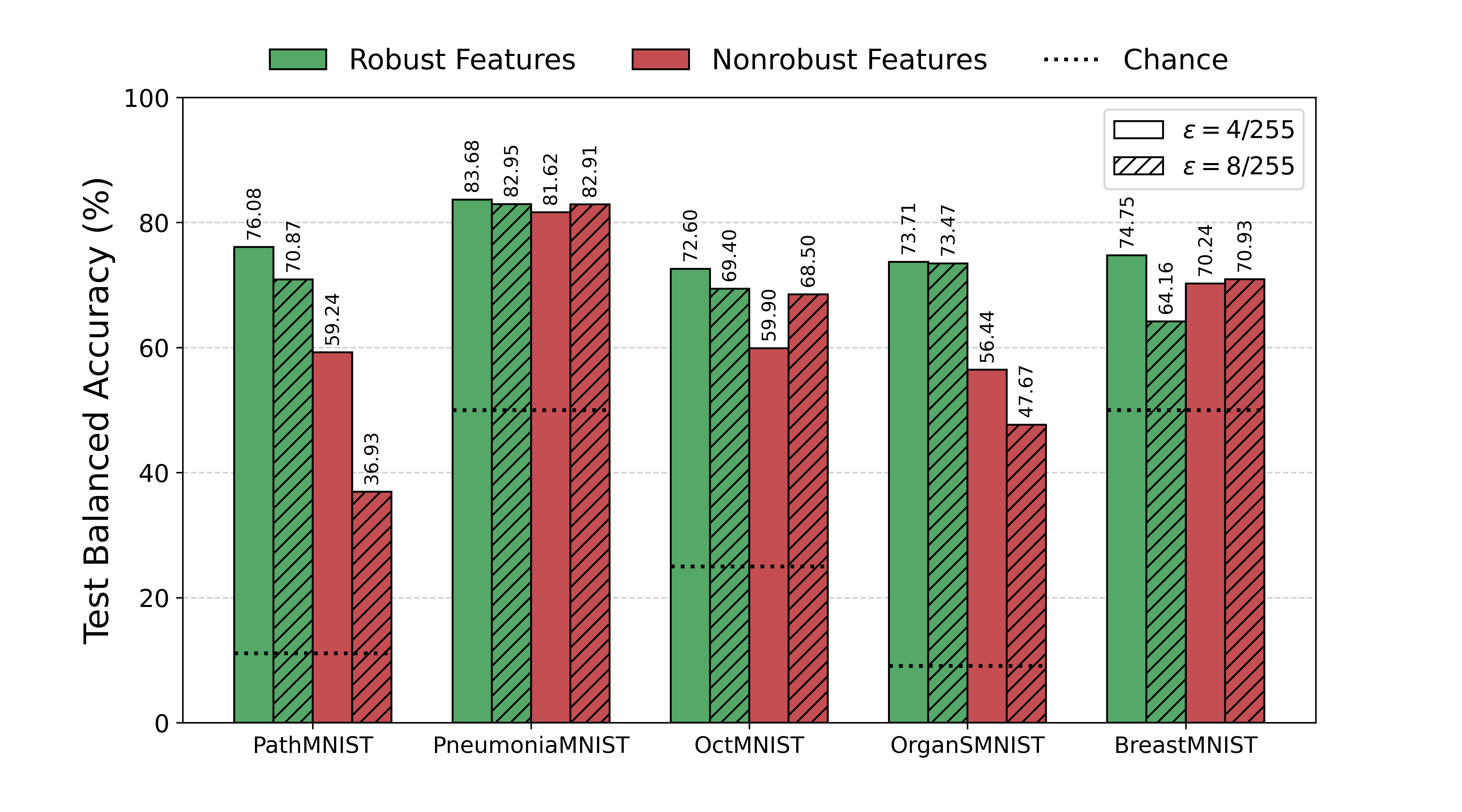}
  \caption{Test 
  accuracy comparison of robust and nonrobust models trained with different values of $\epsilon$. Solid bars: $\epsilon=\frac{4}{255}$. Dashed bars: $\epsilon=\frac{8}{255}$}
  \label{fig:epsilon_comparison}
\end{figure}

Fig.~\ref{fig:epsilon_comparison} compares the in-distribution performance of robust and nonrobust models trained using $\epsilon=\frac{4}{255}$ and $\epsilon=\frac{8}{255}$. Firstly, we observe that robust models trained with the larger perturbation budget consistently exhibit slightly lower test performance. This is expected, as larger perturbations impose a stronger constraint on the features that can be reliably used for classification by the model. Secondly, the performance of the nonrobust-only models varies across datasets, with slight improvements in some cases and larger decreases in others.
Overall, the trends remain unchanged: robust models still trade off standard accuracy for improved robustness, while nonrobust-only models continue to achieve well above chance performance.
\section{Hyperparameter details}
\label{sec:app_hp}

In this section we explain the hyperparameter search methodology used to train the models presented in the main paper. For all models we make use of a WRN-16-8 architecture with swish activations, a cosine annealing learning rate scheduler, and SGD with $0.9$ Nesterov momentum. 

For the standard and nonrobust-only models, we perform a grid search over learning rate, batch size, and one additional hyperparameter: data augmentation for nonrobust-only models, and maximum number of epochs for standard models. Varying the number of epochs effectively changes the learning rate schedule under cosine annealing. Also note that we use smaller batch size values for BreastMNIST and PneumoniaMNIST, as these datasets have substantially smaller training sets. This grid search results in a total of 32 runs per dataset. Model selection is then performed by comparing validation set performance, ensuring that no tuning is performed on the test set.

For the robust models that are adversarially trained, we perform the same grid search as used for the standard models, but reduce the search space for some datasets due to the significantly higher computational cost of adversarial training (approximately 10× when using 10-step PGD). Depending on the dataset, between 12 and 32 configurations are evaluated.

\begin{table}[h]
\centering
\caption{Hyperparameter search space for standard and nonrobust-only models.}
\label{tab:base_hparams}
\begin{tabular}{lll}
\toprule
Hyperparameter & Values & Notes \\
\midrule
Learning rate & $\{0.4, 0.3, 0.2, 0.1\}$ & \\
Batch size & $\{1024, 512, 256, 32\}$ & For large datasets \\
            & $\{256, 128, 64, 32\}$   & For small datasets \\
Data augmentation & \{True, False\} & For nonrobust-only models \\
Max epochs & $\{200, 400\}$ & For standard models \\
\bottomrule
\end{tabular}
\end{table}

\end{document}